\documentclass{ws-procs11x85}
\usepackage{ws-procs-thm}           

\usepackage{amsmath,amssymb}
\usepackage{bbm}
\usepackage{dsfont}
\usepackage{bbold}
\usepackage{bm}
\usepackage{tabularx}

\usepackage{wrapfig}

\makeatletter
\newcommand\notsotiny{\@setfontsize\notsotiny{7.5}{8.5}}

\DeclareMathOperator*{\E}{\mathbb{E}}

\usepackage[algo2e]{algorithm2e}
\newsavebox{\negativesamplingloss}

\usepackage[hidelinks]{hyperref}

\begin{document}

\title{From genome to phenome: Predicting multiple cancer phenotypes based on somatic genomic alterations via the genomic impact transformer}

\author{Yifeng Tao$^1$, Chunhui Cai$^2$, William W. Cohen$^{1, \dag}$ and Xinghua Lu$^{2, 3, \dag}$}

\address{
	$^1$School of Computer Science, Carnegie Mellon University,\\
	$^2$Department of Biomedical Informatics, University of Pittsburgh, \\
	$^3$Department of Pharmaceutical Sciences, School of Pharmacy, University of Pittsburgh, \\
Pittsburgh, PA, USA\\
$^\dag$To whom correspondence should be addressed. E-mail: wcohen@cs.cmu.edu, xinghua@pitt.edu}

%

\begin{abstract}
Cancers are mainly caused by somatic genomic alterations (SGAs) that perturb cellular signaling systems and eventually activate oncogenic processes. Therefore, understanding the functional impact of SGAs is a fundamental task in cancer biology and precision oncology.
Here, we present a deep neural network model with encoder-decoder architecture, referred to as genomic impact transformer (GIT), to infer the functional impact of SGAs on cellular signaling systems through modeling the statistical relationships between SGA events and differentially expressed genes (DEGs) in tumors. The model utilizes a multi-head self-attention mechanism to identify SGAs that likely cause DEGs, or in other words, differentiating potential driver SGAs from passenger ones in a tumor. GIT model learns a vector (gene embedding) as an abstract representation of functional impact for each SGA-affected gene. Given SGAs of a tumor, the model can instantiate the states of the hidden layer, providing an abstract representation (tumor embedding) reflecting characteristics of perturbed molecular/cellular processes in the tumor, which in turn can be used to predict multiple phenotypes.
We apply the GIT model to 4,468 tumors profiled by The Cancer Genome Atlas (TCGA) project. The attention mechanism enables the model to better capture the statistical relationship between SGAs and DEGs than conventional methods, and distinguishes cancer drivers from passengers. The learned gene embeddings capture the functional similarity of SGAs perturbing common pathways. The tumor embeddings are shown to be useful for tumor status representation, and phenotype prediction including patient survival time and drug response of cancer cell lines.\footnote{Supplemental information (SI), GIT model, pre-processed TCGA data, and gene embeddings are available at \href{https://github.com/yifengtao/genome-transformer}{https://github.com/yifengtao/genome-transformer}.}
\end{abstract}


\copyrightinfo{\copyright\ 2019 The Authors. Open Access chapter published by World Scientific Publishing Company and distributed under the terms of the Creative Commons Attribution Non-Commercial (CC BY-NC) 4.0 License.}

\section{Introduction}

Cancer is mainly caused by the activation of oncogenes or deactivation of tumor suppressor genes (collectively called ``driver genes'') as results of somatic genomic alterations (SGAs)\cite{sga13b}, including  somatic mutations (SMs)\cite{sna13, sna14}, somatic copy number alterations (SCNAs)\cite{cna13a, cna13b}, DNA structure variations (SVs)\cite{sv14a}, and epigenetic changes\cite{epg12a}.
Precision oncology relies on the capability of identifying and targeting tumor-specific aberrations resulting from driver SGAs and their effects on molecular and cellular phenotypes. However, our knowledge of driver SGAs and cancer pathways remains incomplete. Particularly, it remains a challenge to determine which SGAs (among often  hundreds) in a specific tumor are drivers, which cellular signals or biological processes a driver SGA perturbs, and which molecular/cellular phenotypes a driver SGA affects.

Current methods for identifying driver genes mainly concentrate on identifying genes that are mutated at a frequency above expectation, based on the assumption that mutations in these genes may provide oncogenic advantages and thus are positively selected\cite{dvfreq12, dvfreq13}. Some works further focus on the mutations perturbing conserved (potentially functional) domains of proteins as indications they may be driver events\cite{dvprot11, dvprot16}. However, these methods do not provide any information regarding the functional impact of enriched mutations on molecular/cellular phenotypes of cells. Without the knowledge of functional impact, it is difficult to further determine whether an SGA will lead to specific molecular, cellular and clinical phenotypes, such as response to therapies. What's more, while both SMs and SCNAs may activate/deactivate a driver gene, there is no well-established frequency-based method that combines different types of SGAs to determine their functional impact.

Conventionally, an SGA event perturbing a gene in a tumor is represented as a ``one-hot'' vector spanning gene space, in which the element corresponding to the perturbed gene is set to ``1''. This representation simply indicates which gene is perturbed, but it does not reflect the functional impact of the SGA, nor can it represent the similarity of distinct SGAs that perturb a common signaling pathway. We conjecture that it is possible to represent an SGA as a low-dimensional vector, in the same manner as the ``word embedding''\cite{word2vec13a, glove14, lstmcrf18} in the natural language processing (NLP) field, such that the representation reflects the functional impact of a gene on biological systems, and genes sharing similar functions should be closely located in such embedding space. Here the ``similar function'' is broadly defined, e.g., genes from the same pathway or of the same biological process\cite{go00}.  
Motivated by this, we propose a scheme for learning ``gene embeddings'' for SGA-affected genes, i.e., a mapping from individual genes to low-dimensional vectors of real numbers that are useful in multiple prediction tasks. 

Based on the assumption that SGAs perturbing cellular signaling systems often eventually lead to changes in gene expression\cite{tci18b}, we introduce an encoder-decoder architecture neural network model called ``genomic impact transformer'' (GIT) to predict DEGs and detect potential cancer drivers with the supervision of DEGs. While deep learning models are being increasingly used to model different bioinformatics problems\cite{deepbio16, deepbio18}, to our knowledge there are few studies using the neural network to model the relationships between SGAs and molecular/cellular phenotypes in cancers. The proposed GIT model has the following innovative characteristics:
(1) The encoder part of the transformer\cite{attn17} first uses SGAs observed in a tumor as inputs, maps each SGA into a gene embedding representation, and combines gene embeddings of SGAs to derive a personalized  ``tumor embedding''. Then the decoder part decodes and translates the tumor embedding to DEGs.
(2) A multi-head self-attention mechanism\cite{attn14, attn15} is utilized in the encoder, which is a technique widely used in NLP to choose the input features that significantly influence the output. 
It differentiates SGAs by assigning different weights to them so that it can potentially distinguish SGAs that have an impact on DEG from those do not, i.e., detecting drivers from passengers.
(3) Pooling inferred weighted impact of SGAs in a tumor produces a personalized tumor embedding, which can be used as an effective feature to predict DEGs and other phenotypes.
(4) Gene embeddings are pre-trained by a ``Gene2Vec'' algorithm and further refined by the GIT, which captures the functional impact of SGAs on the cellular signaling system.
Our results and analysis indicate that above innovative approaches enable us to derive powerful gene embedding and tumor embedding representations that are highly informative of molecular, cellular and clinical phenotypes.

\section{Materials and methods}
\subsection{SGAs and DEGs pre-processing}

We obtained SGA data, including SMs and SCNAs, and DEGs of 4,468 tumors  consisting of 16 cancer types directly from TCGA portal\cite{tcga13}. Details available in SI (\sref{sec:preprocess}).

\subsection{The GIT neural network}

\label{sec:netstruct}

\begin{figure}[!tbp] 
	\centerline{\includegraphics[width=0.75\linewidth]{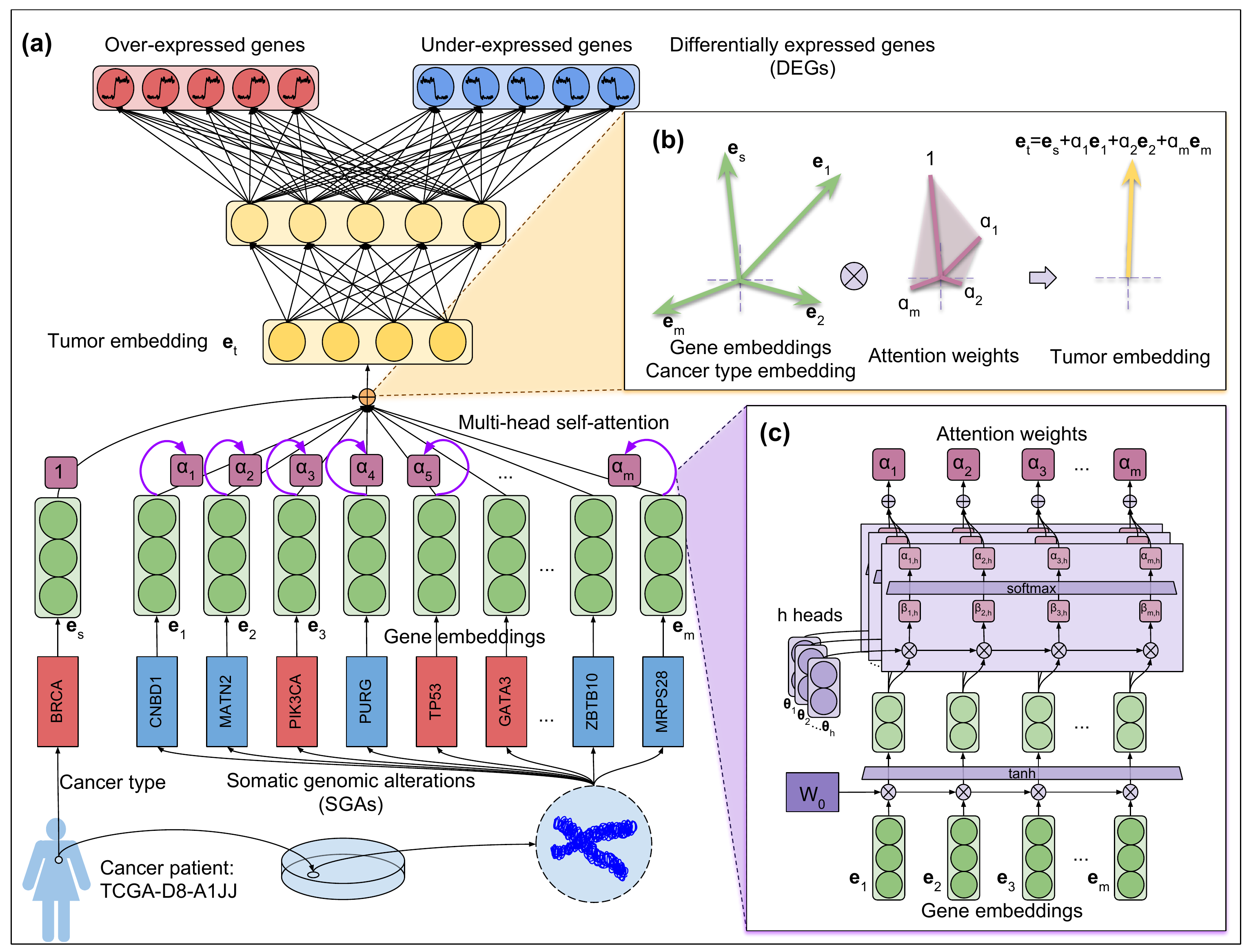}}
	\caption{
		\textbf{(a)} Overall architecture of GIT. An example case and its detected drivers are shown.
		\textbf{(b)} A two-dimensional demo that shows how attention mechanism combines multiple gene embeddings of SGAs $\left\{\mathbf{e}_g\right\}_{g=1}^m$ and cancer type embedding $\mathbf{e}_s$ into a tumor embedding vector $\mathbf{e}_t$ using attention weights $\left\{ \alpha_g \right\}_{g=1}^m$.
		\textbf{(c)} Calculation of attention weights  $\left\{ \alpha_g \right\}_{g=1}^m$ using gene embeddings $\left\{\mathbf{e}_g\right\}_{g=1}^m$. 
	}
	\label{fig:structure}
\end{figure}

\subsubsection{GIT network structure: encoder-decoder architecture}

\Fref{fig:structure}a shows the general structure of the GIT model with an overall encoder-decoder architecture. GIT mimics hierarchically organized cellular signaling system\cite{chen2015, chen2016}, in which a neuron may potentially encode the signal of one or more signaling proteins. When a cellular signaling system is perturbed by SGAs, it often can lead to changes in measured molecular phenotypes, such as gene expression changes.
Thus, for a tumor $t$, the set of its SGAs $\left\{ g \right\}_{g=1}^m$ is connected to the GIT neural network as observed input (\fref{fig:structure}a bottom part squares). The impact of SGAs is represented as embedding vectors $\left\{ \mathbf{e}_g \right\}_{g=1}^m$, which are further linearly combined to produce a tumor embedding vector $\mathbf{e}_t$ through an attention mechanism in the encoder  (\fref{fig:structure}a middle part).
We explicitly represent cancer type $s$ and its influence on encoding system  $\mathbf{e}_s$ of the tumor because tissue type influences which genes are expressed in cells of specific tissue as well. Finally, the decoder module, which consists of a feed-forward multi-layer perceptron (MLP)\cite{mlp58}, transforms the functional impact of SGAs and cancer type into DEGs of the tumor (\fref{fig:structure}a top part).

\subsubsection{Pre-training gene embeddings using Gene2Vec algorithm}

In this study, we projected the discrete binary representation of SGAs perturbing a gene into a continuous embedding space, which we call ``gene embeddings'' of corresponding SGAs, using a ``Gene2Vec'' algorithm, based on the assumption of co-occurrence pattern of SGAs in each tumor, including mutually exclusive patterns of mutations affecting a common pathway\cite{mutationpattern12}. 
These gene embeddings were further updated and fine-tuned by the GIT model with the supervision of affected DEGs. Algorithm details available in SI (\sref{sec:gene2vec}).

\subsubsection{Encoder: multi-head self-attention mechanism}

\label{sec:attention}

To detect the difference of functional impact of SGAs in a tumor, we designed a multi-head self-attention mechanism (\fref{fig:structure}a middle part). For all SGA-affected genes $\left\{ g \right\}_{g=1}^m$ and the cancer type $s$ of a tumor $t$, we first mapped them to corresponding gene embeddings $\left\{ \mathbf{e}_g \right\}_{g=1}^m$ and a cancer type embedding $\mathbf{e}_s$ from a look-up table $\mathcal{E}\!=\!\left\{ \mathbf{e}_g\right\}_{g \in \mathcal{G}} \cap \left\{ \mathbf{e}_s \right\}_{s \in \mathcal{S}}$, where $\mathbf{e}_g$ and $\mathbf{e}_s$ are  real-valued vectors. From the implementation perspective, we treated cancer types in the same way as SGAs, except the attention weight of it is fixed to be ``1''.
The overall idea of producing the tumor embedding $\mathbf{e}_t$ is to use the weighted sum of cancer type embedding $\mathbf{e}_s$ and gene embeddings $\left\{ \mathbf{e}_g \right\}_{g=1}^m$ (\fref{fig:structure}b) :
\begin{equation}
	\mathbf{e}_t = 1 \cdot \mathbf{e}_s + \sum\nolimits_{g} \alpha_g \cdot \mathbf{e}_g = 1 \cdot \mathbf{e}_s + \alpha_1 \cdot \mathbf{e}_1 + ... + \alpha_m \cdot \mathbf{e}_m.
	\label{eq:tumor-embedding}
\end{equation}
The attention weights $\left\{ \alpha_{g} \right\}_{g=1}^m$ were calculated by employing multi-head self-attention mechanism, using gene embeddings of SGAs $\left\{ \mathbf{e}_g \right\}_{g=1}^m$ in the tumor: $\left\{ \alpha_{g} \right\}_{g=1}^m = \text{Function}_{\text{Attention}}\left(\left\{ \mathbf{e}_g \right\}_{g=1}^m ; W_0, \Theta \right)$ (\fref{fig:structure}c). See SI (\sref{sec:mathmatics}) for mathematical details.
Overall we have three parameters $\left\{ W_0, \Theta, \mathcal{E} \right\}$ to train in the multi-head attention module using back-propagation\cite{backprop86}. The look-up table $\left\{ \mathbf{e}_g \right\}_{g\in\mathcal{G}}$ was initialized with Gene2Vec pre-trained gene embeddings and refined by GIT here.

\subsubsection{Decoder: multi-layer perceptron (MLP)}

For a specific tumor $t$, we fed tumor embedding $\mathbf{e}_t$ into an MLP with one hidden layer as the decoder, using non-linear activation functions and fully connected layers, to produce the final predictions $\hat{y}$ for DEGs $y$; (\fref{fig:structure}a top part):
\begin{equation}
\hat{y} = \sigma( W_{2} \cdot \text{ReLU} ( W_{1} \cdot \text{ReLU} (\mathbf{e}_t) + b_{1}  )   + b_{2} ).
\label{eq:transformer-mlp}
\end{equation}
where $\text{ReLU}(x)\!=\!\max(0, x)$ is rectified linear unit, and $\sigma(x)\!=\!\left(1\!+\!\exp(-x) \right)^{-1}$ is sigmoid activation function.
The output of the decoder and actual values of DEGs were used to calculate the $\ell_2$-regularized cross entropy, which was minimized during training: $\min_{\mathcal{W}, \mathcal{E}, \Theta, b}\text{CrossEnt}(y, \hat{y}) + \ell_2(\mathcal{W}, \mathcal{E}, \Theta ; \lambda_2)$,
where $\mathcal{W} = \left\{ W_l \right\}_{l=0}^2$, cross entropy loss defined as $\text{CrossEnt} (y, \hat{y}) = -\sum\nolimits_i \left[ (1-y_i) \log(1-\hat{y}_{i}) + y_i \log \hat{y}_{i} \right]$, and $\ell_p$ regularizer defined as $\ell_p(\mathcal{W} ; \lambda) = \lambda \cdot \sum\nolimits_l \left\lVert W_l \right\lVert_p, p \in \{1, 2\}$.

\subsection{Training and evaluation}

We utilized PyTorch (https://pytorch.org/) to train, validate and test the Gene2Vec, GIT (variants) and other conventional models (Lasso and MLPs; Section~\ref{sec:statistical}).
The training, validation and test sets were split in the ratio of 0.33:0.33:0.33 and fixed across different models. The hyperparameters were tuned over the training and validation sets to get best F1 scores, trained on training and validation sets, and finally applied to the test set for evaluation if not further mentioned below.
The models were trained by updating parameters using backpropagation\cite{backprop86}, specifically, using mini-batch Adam\cite{adam15} with default momentum parameters. Gene2Vec used mini-batch stochastic gradient descent (SGD) instead of Adam. Dropout\cite{dropout12} and weight decay ($\ell_p$-regularization) were used to prevent overfitting. We trained all the models 30 to 42 epochs until they fully converged.
The output DEGs were represented as a sparse binary vector.  We utilized various performance metrics including accuracy, precision, recall, and F1 score, where F1 is the harmonic mean of precision and recall. The training and test were repeated for five runs get the mean and variance of evaluation metrics. We designed two metrics in the present work for evaluating the functional similarity among genes sharing similar gene embedding: ``nearest neighborhood (NN) accuracy'' and ``GO enrichment''. See SI (\sref{sec:evaluationembedding}) for the definition and meaning of them.

\section{Results}
\subsection{GIT statistically detects real biological signals}

\label{sec:statistical}

\begin{wraptable}{l}{0.60\textwidth}
	\tbl{Performances of GIT (variants) and baseline methods.}
	{\begin{tabular}{ m{0.22\linewidth} m{0.16\linewidth} m{0.16\linewidth}  m{0.16\linewidth} m{0.16\linewidth} }
			\hline
			Methods & Precision & Recall & F1 score & Accuracy \\
			\hline
			Lasso & 59.6$\pm$0.05 & 52.8$\pm$0.03 & 56.0$\pm$0.01 & 74.0$\pm$0.02 \\ 
			1 layer MLP & 61.9$\pm$0.09 & 50.4$\pm$0.17 & 55.6$\pm$0.07 & 74.7$\pm$0.02 \\ 
			2 layer MLP & 64.2$\pm$0.39 & 52.0$\pm$0.66 & 57.4$\pm$0.28 & 75.9$\pm$0.09 \\ 
			3 layer MLP & 64.2$\pm$0.37 & 50.5$\pm$0.30 & 56.5$\pm$0.19 & 75.7$\pm$0.13 \\
			\hline
			GIT - can & 60.5$\pm$0.34 & 45.8$\pm$0.38 & 52.1$\pm$0.29 & 73.6$\pm$0.14 \\ 
			GIT - attn & 67.6$\pm$0.32 & 55.3$\pm$0.77& 60.8$\pm$0.35 & 77.7$\pm$0.05 \\ 
			GIT - init & \textbf{69.8}$\pm$0.28 & 54.1$\pm$0.37 & 60.9$\pm$0.16 &  78.3$\pm$0.06 \\
			\hline 
			GIT & 69.5$\pm$0.09 & \textbf{57.1}$\pm$0.18 & \textbf{62.7}$\pm$0.08 & \textbf{78.7}$\pm$0.01  \\
			\hline
	\end{tabular}}\label{tab:performance}
\end{wraptable}

The task of GIT is to predict DEGs (dependent variables) using SGAs as input (independent variables). 
Our results of GIT performance on both real and shuffled data demonstrates that GIT is able to capture real statistical relationships between SGAs and DEGs from the noisy biological data (SI: Sec.~\ref{sec:shuffle}).

As a comparison, we also trained and tested the Lasso (multivariate regression with $\ell_1$-regularization)\cite{lasso96} and MLPs\cite{mlp58} as baseline prediction models to predict DEGs based on SGAs. 
The Lasso model is appealing in our setting because, when predicting a DEG, it can filter out most of the irrelevant input variables (SGAs) and keep only the most informative ones, and it is a natural choice in our case where there are 19.8k possible SGAs.
However, in comparison to MLP, it lacks the capability of portraying complex relationships between SGAs and DEGs. On the other hand, while conventional MLPs have sufficient power to capture complex relationships--particularly, the neurons in hidden layers may mimic signaling proteins\cite{chen2016}--they can not utilize any biological knowledge extracted from cancer genomics, nor do they explain the signaling process and distinguish driver SGAs.
We employed the precision, recall, F1 score, as well as accuracy to compare GIT and traditional methods (Table~\ref{tab:performance}: 1st to 4th, and last rows). One can conclude that GIT outperforms all these other conventional baseline methods for predicting DEGs in all metrics, indicating the specifically designed structure of GIT is able to soar the performance in the task of predicting DEGs from SGAs.

In order to evaluate the utility of each module (procedure) in GIT, we conducted ablation study by removing one module at a time: the cancer type input (``can''), the multi-head self-attention module (``attn''), and the initialization with pre-trained gene embeddings (``init''). The impact of each module can be detected by comparing to the full GIT model. All the modules in GIT help to improve the prediction of DEGs from SGAs in terms of overall performance: F1 score and accuracy (Table~\ref{tab:performance}: 5th to last rows).

\begin{figure*}[!tbp]
	\centerline{\includegraphics[width=0.90 \linewidth]{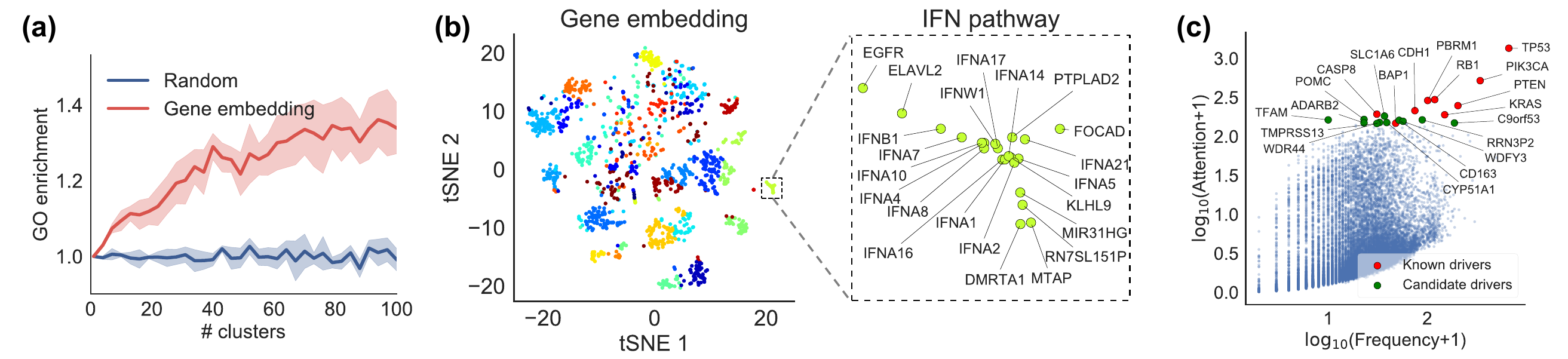}}
	\caption{
		\textbf{(a)} GO enrichment of vs. number of groups in \textit{k}-means clustering.
		\textbf{(b)} t-SNE visualization of gene embeddings. The different colors represent \textit{k}-means (40 clusters) clustering labels. An enlarged inset of a cluster is shown, which contains a set of closely related genes which we refer to ``IFN pathway''.
		\textbf{(c)} Landscape of attention of SGAs based on attention weights and frequencies.
	}
	\label{fig:results}
\end{figure*}

\subsection{Gene embeddings compactly represent the functional impact of SGAs}

\label{sec:sga}

We examined whether the gene embeddings capture the functional similarity of SGAs, using mainly two metrics: NN accuracy and GO enrichment (Defined in SI \sref{sec:evaluationembedding}).
\textbf{NN accuracy:} By capturing the co-occurrence pattern of somatic alterations, the Gene2Vec pre-trained gene embeddings improve 36\% in NN accuracy over the random chance of any pair of the genes sharing Gene Ontology (GO) annotation\cite{go00} (\tref{tab:nnaccuracy}). The fine-tuned embeddings by GIT further show a one-fold increase in NN accuracy. These results indicate that the learned gene embeddings are consistent with the gene functions, and they map the discrete binary SGA representation into a meaningful and compact space.
\textbf{GO enrichment:} We performed clustering analysis of SGAs in embedding space using \textit{k}-means clustering, and calculated GO enrichment, and we varied the number of clusters (\textit{k}) to derive clusters with different degrees of granularity (Fig.~\ref{fig:results}a).
As one can see, when the genes are randomly distributed in the embedding space, they get GO enrichment of 1. However, in the gene embedding space, the GO enrichment increases fast until the number of clusters reaches 40, indicating a strong correlation between the clusters in embedding space and the functions of the genes.

\begin{wraptable}{lht}{0.45\textwidth}
	\tbl{NN accuracy with respect to GO in different gene embedding spaces.}
	{\begin{tabular}{ m{0.40\linewidth} m{0.3\linewidth} m{0.28\linewidth} }
			\hline
			Gene embeddings & NN accuracy  & Improvement  \\
			\hline
			Random pairs & 5.3$\pm$0.36  & -- \\
			Gene2Vec  & 7.2 & 36\% \\
			Gene2Vec + GIT & \textbf{10.7} & 100\% \\
			\hline
	\end{tabular}}\label{tab:nnaccuracy}
\end{wraptable}

To visualize the manifold of gene embeddings, we grouped the genes into 40 clusters, and conducted the t-SNE \cite{tsne08} of genes (Fig.~\ref{fig:results}b left panel).
Using PANTHER GO enrichment analysis\cite{panther13}, 12 out of 40 clusters are shown to be enriched in at least one biological process (SI \sref{sec:clusterfunction}). Most of the gene clusters are well-defined and tight located in the projected t-SNE space.
As a case study, we took a close look at one cluster (Fig.~\ref{fig:results}b right panel), which contains a set of functionally similar genes, such as that code a protein family of type I interferons (IFNs), which are responsible for immune and viral response\cite{ifnar12}.

\subsection{Self-attention reveals impactful SGAs  on cancer cell transcriptome}

\label{sec:driver}

\begin{wraptable}{lht}{0.7\textwidth}
	\tbl{Top five SGA-affected genes ranked according to attention weight.
	}{
		\begin{tabular}{m{0.02\linewidth}
				m{0.11\linewidth}  
				m{0.11\linewidth}  
				m{0.11\linewidth} 
				m{0.11\linewidth} 
				m{0.11\linewidth} 
				m{0.15\linewidth} 
			}
			\hline
			\multicolumn{1}{l}{Rank} & \multicolumn{1}{l}{PANCAN}  & \multicolumn{1}{l}{BRCA} & \multicolumn{1}{l}{HNSC} & \multicolumn{1}{l}{LUAD} & \multicolumn{1}{l}{GBM} & \multicolumn{1}{l}{BLCA}\\
			\hline
			1 &  \textit{\textbf{TP53}}   & \textit{\textbf{TP53}} & \textit{\textbf{TP53}}   & \textit{\textbf{STK11}}  & \textit{\textbf{TP53}}  & \textit{\textbf{TP53}}   \\ 
			2 &  \textit{\textbf{PIK3CA}}   & \textit{\textbf{PIK3CA}}  & \textit{\textbf{CASP8}}  & \textit{\textbf{TP53}}  & \textit{\textbf{PTEN}} & \textit{\textbf{FGFR3}}   \\
			3 &  \textit{\textbf{RB1}}   & \textit{\textbf{CDH1}}  & \textit{\textbf{PIK3CA}}   & \textit{\textbf{KRAS}}   & \textit{C9orf53} & \textit{\textbf{RB1}}   \\ 
			4 &  \textit{\textbf{PBRM1}}    & \textit{\textbf{GATA3}}   & \textit{\textbf{CYLD}}   & \textit{CYLC2}  & \textit{\textbf{RB1}} & \textit{\textbf{HSP90AA1}}    \\ 
			5 &  \textit{\textbf{PTEN}}    & \textit{MED24}  & \textit{RB1} & \textit{\textbf{KEAP1}} & \textit{CHIC2} & \textit{\textbf{STAG2}} \\
			\hline
	\end{tabular}}\label{tab:attention}
\end{wraptable}

While it is widely accepted that cancer is mainly caused by SGAs, but not all SGAs observed in a cancer cell are causative\cite{sga13b}. Previous methods mainly concentrate on searching for SGAs with higher than expected frequency to differentiate candidate drivers SGAs from passenger SGAs. GIT provides a novel perspective to address the problem: identifying the SGAs that have a functional impact on cellular signaling systems and eventually lead DEGs as the \textit{tumor-specific} candidate drivers. Here we compare the relationship of overall attention weights (inferred by GIT model) and the frequencies of somatic alterations (used as the benchmark/control group) in all the cancer types (Pan-Cancer) from our test data (Fig.~\ref{fig:results}c). In general, the attention weights are  correlated with the alteration frequencies of genes, e.g., common cancer drivers such as \textit{TP53} and \textit{PIK3CA} are the top two SGAs selected by both methods\cite{sna13}. However, our self-attention mechanism assigns high weights to many of genes previously not designated as drivers, indicating these genes are potential cancer drivers although their roles in cancer development remain to be further studied. Table~\ref{tab:attention} lists top SGAs ranked according to GIT attention weights in pan-cancer and five selected cancer types, where known cancer drivers from TumorPortal\cite{sna14} and IntOGen\cite{intogen13} are marked as bold font.
Apart from \textit{TP53} and \textit{PIK3CA} as drivers in the pan-cancer analysis\cite{sna13}, we also find the top cancer drivers in specific cancer types consistent with our knowledge of cancer oncology. For example, \textit{CDH1} and \textit{GATA3} are drivers of breast invasive carcinoma (BRCA)\cite{tcga12}, \textit{CASP8} is known driver of head and neck squamous cell carcinoma (HNSC)\cite{hnsc11}, \textit{STK11}, \textit{KRAS}, \textit{KEAP1} are known drivers of lung adenocarcinoma (LUAD)\cite{luad14}, \textit{PTEN} and \textit{RB1} are drivers of glioblastoma (GBM)\cite{gbm13}, and \textit{FGFR3}, \textit{RB1}, \textit{HSP90AA1}, \textit{STAG2} are known drivers in urothelial bladder carcinoma (BLCA)\cite{blca14}. 
In contrast, the most frequently mutated genes (control group) are quite different from that using attention mechanism (experiment group), and only a few of them are known drivers (SI \sref{sec:rank}).

\subsection{Personalized tumor embeddings reveal distinct survival profiles}

\label{sec:survival}

\begin{figure}[!t]
	\centerline{\includegraphics[width=0.90 \linewidth]{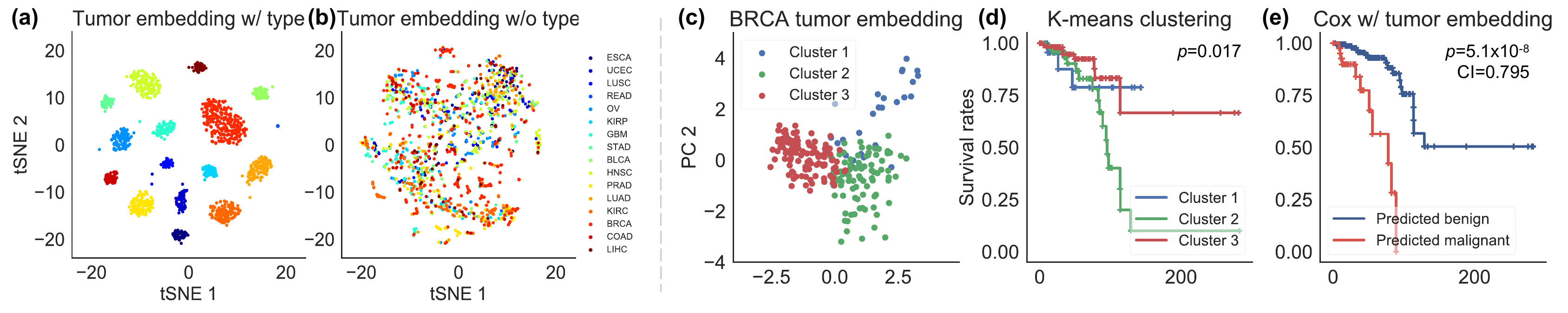}}
	\caption{
		\textbf{(a)} t-SNE of full tumor embedding $\mathbf{e}_t$. 
		\textbf{(b)} t-SNE of stratified tumor embedding ($\mathbf{e}_t$-$\mathbf{e}_s$). 
		\textbf{(c)} PCA of tumor embedding shows internal subtype structure of BRCA tumors. Color lablels the group index of \textit{k}-means clustering. 
		\textbf{(d)} KM estimators of the three breast cancer groups.
		\textbf{(e)} Cox regression using tumor embeddings.
	}
	\label{fig:survival}
\end{figure}

Besides learning the specific biological function impact of SGAs on DEGs, we further examined the utility of tumor embeddings $\mathbf{e}_t$ in two perspectives: (1) Discovering patterns of tumors potentially sharing common disease mechanisms across different cancer types; (2) Using tumor embedding to predict patient survival.

We first used the t-SNE plot of tumor embeddings to illustrate the common disease mechanisms across different cancer types (Fig.~\ref{fig:survival}a). When cancer type embedding $\mathbf{e}_s$ is included in full tumor embedding $\mathbf{e}_t$, which has a much higher weight than any individual gene embedding (Fig.~\ref{fig:structure}b, Eq.~\ref{eq:tumor-embedding}) and dominates the full tumor embedding, tumor samples are clustered according to cancer types. This is not surprising as it is well appreciated that expressions of many genes are tissue-specific\cite{cancertissue14}.
To examine the pure effect of SGAs on tumor embedding, we removed the effect of tissue by subtracting cancer type embeddings $\mathbf{e}_s$, followed by clustering tumors in the stratified tumor embedding space (Fig.~\ref{fig:survival}b). 
It is interesting to see that each dense area (potential tumor clusters) includes tumors from different tissues of origins, indicating SGAs in these tumors may reflect shared disease mechanisms (pathway perturbations) among tumors, warranting further investigations.

The second set of experiments was to test whether differences in tumor embeddings (thereby difference in disease mechanisms) are predictive of patient clinical outcomes. We conducted unsupervised \textit{k}-means clustering using only breast cancer tumors from our test set, which reveals 3 three groups (\fref{fig:survival}c) with significant difference in survival profiles evaluated by log-rank test\cite{logrank66} (\fref{fig:survival}d; \textit{p}-value=0.017). 
In addition, using tumor embeddings as input features, we trained $\ell_{1,2}$-regularized (elastic net)\cite{elasticnet05} Cox proportional hazard models\cite{coxreg92} in a 10-fold cross-validation (CV) experiment. This led to an informative ranked list of tumors according to predicted survivals/hazards evaluated by the concordance index (CI) value (CI=0.795), indicating that the trained model is very accurate. We further split test samples into two groups divided by the median of predicted survivals/hazards, which also yields significant separation of patients in survival profiles (\fref{fig:survival}e; \textit{p}-value=$5.1\times10^{-8}$), indicating that our algorithm has correctly ranked the patients according to characteristics of the tumor.

As shown above, distinct SGAs may share similar embeddings if they share similar functional impact. Thus, two tumors may have similar tumor embeddings even though they do not share any SGAs, as long as the functional impact of distinct SGAs from these tumors are similar. Therefore,  tumor embedding makes it easier to discover common disease mechanisms and their impact on patient survival.
To further test this, we also performed clustering analysis on breast cancer tumors represented in original SGA space, followed similar survival analysis as described in the previous paragraph (SI \sref{sec:sgasurvival}).

\subsection{Tumor embeddings are predictive of drug responses of cancer cell lines }

\label{sec:drug}

\begin{wrapfigure}{l}{0.32\textwidth}
	\begin{center}
		\includegraphics[width=1.0\linewidth]{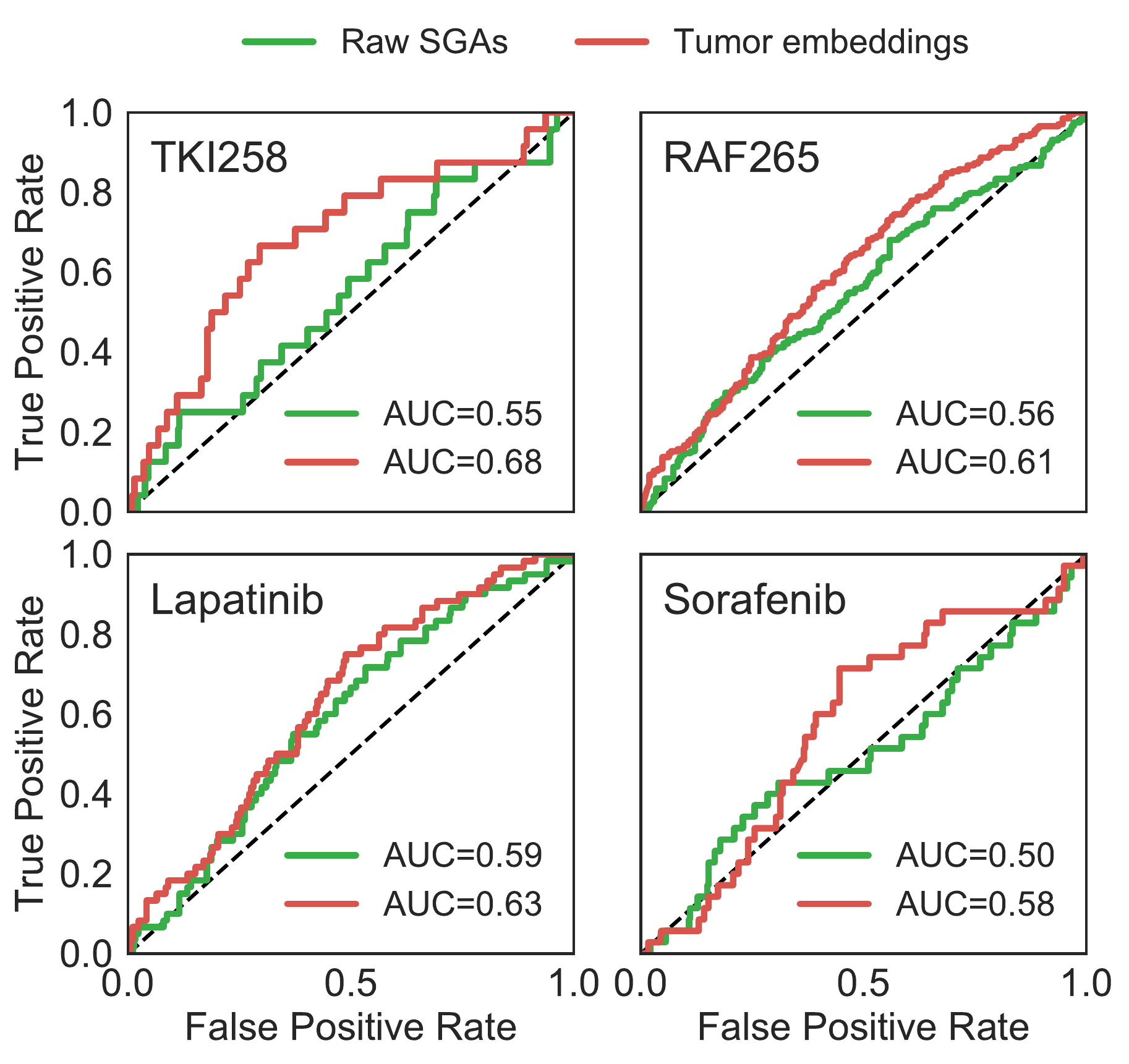}
	\end{center}
	\caption{
		ROC curves and the areas under the curve (AUCs) of Lasso models trained with original SGAs and tumor embeddings representations on predicting responses to four drugs.
	}\label{fig:drug}
\end{wrapfigure}

Precision oncology concentrates on using patient-specific omics data to determine optimal therapies for a patient.  We set out to see if SGA data of cancer cells can be used to predict their sensitivity to anti-cancer drugs. We used the CCLE dataset\cite{ccle12}, which performed drug sensitivity screening over hundreds of cancer cell lines and 24 anti-cancer drugs. The study collects genomic and transcriptomic data of these cell lines, but in general, the genomic data (except the molecularly targeted genes) from a cell line are not sufficient to predict sensitivity its sensitivity to different drugs.

We discretized the response of each drug following the procedure in previous research\cite{ding17, ccle12}.  
Since CCLE only contains a small subset of mutations in TCGA dataset (around 1,600 gene mutations), we retrained the GIT with this limited set of SGAs in TCGA, using default hyperparameters we set before. Cancer type input was removed as well, which is not explicitly provided in CCLE dataset.
The output of tumor embeddings $\mathbf{e}_t$ was then extracted as feature.
We formulated drug response prediction as a binary classification problem with $\ell_{1}$-regularized cross entropy loss (Lasso), where the input can be raw sparse SGAs or tanh-curved tumor embeddings $\text{tanh}(\mathbf{e}_t)$. 
Following previous work\cite{ccle12}, we performed 10-fold CV experiment training Lasso using either inputs to test the drug response prediction task of four drugs with distinct targets.
Lasso regression using tumor embeddings consistently outperforms the models trained with original SGAs as inputs (Fig.~\ref{fig:drug}).  Specifically, in the case of Sorafenib, the raw mutations just give random prediction results, while the tumor embedding is able to give predictable results.
It should be noted that it is possible that certain cancer cells may host SGAs along the pathways related to FGFR, RAF, EGFR, and RTK, rendering them sensitive to the above drugs. Such information can be implicitly captured and represented by the tumor embeddings, so that the information from raw SGAs are captured and pooled to enhance classification accuracy.

\section{Conclusion and Future Work}

Despite the significant advances in cancer biology, it remains a challenge to reveal disease mechanisms of each individual tumor, particularly which and how SGAs in a cancer cell lead to the development of cancer.
Here we propose the GIT model to learn the general impact of SGAs, in the form of gene embeddings, and to precisely portray their effects on the downstream DEGs with higher accuracy. With the supervision of DEGs, we can further assess the importance of an SGA using multi-head self-attention mechanisms in each individual tumor. More importantly, while the tumor embeddings are trained with predicting DEGs as the task, it contains information for predicting other phenotypes of cancer cells, such as patient survival and cancer cell drug sensitivity.
The key advantage of transforming SGA into a gene embedding space is that it enables the detection and representation of the functional impact of SGAs on cellular processes, which in turn enables detection of common disease mechanisms of tumors even if they host different SGAs. 
We anticipate that GIT, or other future models like it, can be applied broadly to gain mechanistic insights of how genomic alterations (or other perturbations) lead to specific phenotypes, thus providing a general tool to connect genome to phenome in different biological fields and genetic diseases. 
One should also be careful that despite the correlation of genomic alterations and phenotypes such as survival profiles and drug response, the model may not fully reveal the causalities and there may exist other confounding factors not considered.

There are a few future directions for further improving the GIT model. 
First of all, decades of biomedical research has accumulated a rich body of knowledge, e.g.,  Gene Ontology and gene regulatory networks, which may be incorporated as the prior of the model to boost the performance\cite{hiearachy18}.
Secondly, we expect that by getting a larger corpus of tumor data with mutations and gene expressions, we will be able to train better models to minimize potential overfitting or variance.
Lastly, more clinically oriented investigations are warranted to examine, when trained with a large volume of tumor omics data, the learned embeddings of SGAs and tumors may be applied to predict sensitivity or resistance to anti-cancer drugs based SGA data that are becoming readily available in contemporary oncology practice.

\section*{Acknowledgments}

We would like to thank Yifan Xue and Michael Q. Ding for providing the processed TCGA data and discretized CCLE data. We also thank to the helpful suggestions from anonymous reviewers.

\section*{Funding}

This work has been partially supported by the following NIH grants: R01LM012011, R01LM010144, and U54HG008540, and it has also been partially supported by the Grant \#4100070287 awarded by the Pennsylvania Department of Health. The content is solely the responsibility of the authors and does not necessarily represent the official views of the above funding agencies.

\bibliographystyle{ws-procs11x85}
\bibliography{main}

\newpage

\setcounter{table}{0}
\renewcommand{\thetable}{S\arabic{table}}

\setcounter{figure}{0}
\renewcommand{\thefigure}{S\arabic{figure}}


\setcounter{equation}{0}
\renewcommand{\theequation}{S\arabic{equation}}

\section*{Supplementary information}
\addcontentsline{toc}{section}{Supplementary information}
\renewcommand{\thesubsection}{S\arabic{subsection}}

\subsection{Data pre-processing of SGAs and DEGs}

\label{sec:preprocess}

We obtained SGA data, including SMs and SCNAs of 4,468 tumors consisting of 16 cancer types\footnote{Instead of single cancer types, we used all the available samples of various cancer types, to find the common signaling mechanisms SGAs in cancer. In addition, the GIT model benefits from the large scale dataset. The heterogeneity of different cancer types was stratified by the additional cancer type feature as input to the model.} directly from TCGA portal\cite{tcga13} and Firehose browser of the Broad Institute\footnote{http://gdac.broadinstitute.org/}.
For SMs: We considered all the non-synonymous mutation events of all genes and considered the mutation events at the gene level, where a mutated gene is defined as one that contains one or more non-synonymous mutations or indels.
For SCNAs: TCGA network discretizes the gene SCNA into 5 different levels: homozygous deletion, single copy deletion, diploid normal copy, low copy number amplification, and high copy number amplification. We only included genes with homozygous deletion (potentially significant loss of gene function) or high copy number amplification (potentially significant gain of gene function) for further analysis, and filtered out the other three types of not-so-significant SCNAs.
Therefore, we collectively designated all SGAs affecting a gene using the name of the gene being perturbed. 
Note that the preprocessing step of SMs and SCNAs excluded the obvious tumor passenger SGAs, since the functions of these mutated genes are not or only slightly perturbed. The remaining SGAs have the potential of being cancer drivers, such as oncogenes with gained functions, or tumor suppressor genes with lost functions.
After processing genomic data from TCGA, we used a binary variable in a ``one-hot'' vector to indicate the genomic status of a gene.  For example, we represented the genomic status of \textit{TP53} as 1, if it is perturbed by one or more of SM/SCNA events in a tumor. 

Gene expression data were pre-processed and obtained from the Firehose browser of the Broad Institute. We determined whether a gene is differentially expressed in a tumor by comparing the gene's expression in the tumor against a distribution of the expression values of the gene in the corresponding tissue-specific ``normal'' or control samples. For a given cancer type, assuming the expression of each gene (log 2 based) follows a Gaussian distribution in control sample, we calculated the \textit{p}-values by determining the probability of observing an expression value from control distribution. Following the practice in previous work\cite{tci18b}, if the \textit{p}-value is equal or smaller than 0.005, the gene is considered as differentially expressed in the corresponding tumor. However, if a DEG is associated with an SCNA event affecting it, we remove it from the DEG list of the tumor.

\clearpage
\subsection{Gene2Vec algorithm implementation}

\label{sec:gene2vec}

While gene embeddings can be directly learned using the GIT model, it has been shown in the field of NLP that the pre-trained word embeddings can significantly improve the performance in other related NLP tasks\cite{word2vec13a, lstmcrf18}.
Such pre-trained word embeddings can capture the knowledge of co-occurrence pattern of the words in languages and exhibit sound semantic properties: words of similar semantic meanings are close in embedding space, e.g., $\mathbf{e}_{\text{``each''}} \approx \mathbf{e}_{\text{``every''}}$.
We therefore propose an algorithm called ``Gene2Vec'' to pre-train the gene embeddings, which is closely related the skip gram word2vec\cite{word2vec13a} pre-training algorithm.
The biology rationale behind Gene2Vec algorithm is that we are able to portrait the co-occurrence pattern of SGAs in each tumor, i.e., mutually exclusive mutations\cite{mutationpattern12}, using gene embeddings and gene context embeddings.

Given the gene embedding $\mathbf{e}_g$ of an SGA-affected gene $g$ and context embedding of any possible SGA-affected gene $c'$: $\mathcal{V}\!=\!\left\{ \mathbf{v}_{c'} \right\}_{c' \in \mathcal{G}}$, where $\mathcal{G}$ is the set of all possible SGA-affected genes, the skip gram paradigm assumes the probability that an alteration in gene $c$ happens together with the alteration in gene $g$ within a tumor with probability:
\begin{equation}
\Pr\left(c \in \text{Context}(g) \ | \ g \right)= \frac{
	\exp \left( \mathbf{e}_g^\intercal \mathbf{v}_c \right)
}{
	\sum\nolimits_{c' \in \mathcal{G}} \exp \left(  \mathbf{e}_g^\intercal \mathbf{v}_{c'} \right)
}.
\label{eq:skip-gram}
\end{equation}

We used the negative sampling (NS) technique to approximately maximize the log-likelihood of skip gram, which would otherwise be computationally expensive to optimize if directly following \eref{eq:skip-gram}. Algorithm~\ref{alg:gene2vec} shows implementation of Gene2Vec.

\begin{algorithm2e}[!htbp]
	\footnotesize
	\begin{lrbox}{\negativesamplingloss}
		\verb!NegativeSamplingLoss!
	\end{lrbox}
	\KwData{
		Genomic alterations in each tumor: $\mathcal{T} = \left\{ T_i\!=\!\{g_{i1}, g_{i2}, ..., g_{i m(i)}\} \right\}_{i = 1, 2, ..., N}$.
	}
	\KwResult{
		Pretrained gene embedding of each gene: \\
		$\mathcal{E} = \{ \mathbf{e}_g \in \mathbb{R}^{n} \}_{\ g \in \mathcal{G}}$. \\
		Context gene embeddings: \\
		$\mathcal{V} = \{ \mathbf{v}_g \in \mathbb{R}^{n} \}_{\ g \in \mathcal{G}}$.
	}
	$f(g) \leftarrow \frac{1}{Z} \sum_{i=1}^{N} \mathbb{1}(g \in T_i), \ g \in \mathcal{G}$\tcp*{Gene frequency}
	$f_n(g) \leftarrow \frac{1}{Z_n} f(g)^{3/4}, \ g \in \mathcal{G}$\tcp*{Normalized frequency}
	$\mathbf{e}_g \sim U\left(-\frac{0.5}{n}, \frac{0.5}{n}\right)^{n}, \ \mathbf{v}_g \leftarrow 0^{n}, \ g \in \mathcal{G}$\tcp*{Initialize gene embeddings and context embeddings}
	\While{not converges}{
		$l \leftarrow 0$\tcp*{Total loss of a mini-batch samples}
		\For{$b = 1, 2, ..., \text{batch\_size}$}{
			$g \sim f$ \tcp*{Sample a gene}
			$g_c \sim \text{Context} (g\;;\; \mathcal{T})$\tcp*{Sample a context gene}
			$g_{nr} \sim f_n, \ r = 1, 2, ..., R$ \tcp*{Sample negative context genes}
			$l \leftarrow l + \text{NSLoss} \left(g, g_c, \{ g_{nr} \}_{r=1}^R\;;\; \mathcal{E}, \mathcal{V}\right)$ \tcp*{Update}
		}
		$(\mathcal{E}, \mathcal{V}) \leftarrow (\mathcal{E}, \mathcal{V}) - \eta \cdot \frac{\partial l}{\partial (\mathcal{E}, \mathcal{V})}$ \tcp*{Gradient descent}
	}
	\SetKwFunction{FContext}{Context}
	\SetKwProg{Pn}{Function}{}{\KwRet{$P_c$}}
	\Pn{\FContext{$g\;;\; \mathcal{T}$}}{
		$P_c \leftarrow U\left( \{ g_c \ | \ g_c \in T_i, g \in T_i \}_{i = 1,2,...,N} \right)$ \tcp*{Uniform distribution on sequence of adjacent mutations}
	}
	\SetKwFunction{FMain}{NSLoss}
	\SetKwProg{Pn}{Function}{}{\KwRet{$l$}}
	\Pn{\FMain{$g, g_c, \{ g_{nr} \}_{r=1}^R\;;\; \mathcal{E}, \mathcal{V}$}}{
		$l \leftarrow \log \sigma \left( \mathbf{e}_g^\intercal \mathbf{v}_{g_c} \right) + \sum_{r=1}^{R} \log \sigma \left( -\mathbf{e}_g^\intercal \mathbf{v}_{g_{nr}} \right) $\tcp*{Negative sampling loss of one sample}
	}
	\caption{
		\textbf{Gene2Vec algorithm to pre-train the gene embeddings using skip gram with negative sampling loss.}
		Given the context information of somatic genomic alterations (SGAs) in each cancer patient, i.e., whether two SGAs happened together in a single tumor, we pre-trained the gene embeddings (and context gene embeddings) using similar techniques to word2vec. Skip gram was used to predict the probability of co-occurred SGAs $c$ given a known SGA $g$, as explained in Equation~\ref{eq:skip-gram}. Negative sampling loss was utilized to accelerate the maximization of log-likelihood in the skip gram assumption. 
		Instead of original mutation frequency $f(g)$, the negative sampling frequency of SGA was sub-sampled by scaling to $f(g)^{3/4}$.
		In practice, the step size $\eta$ in mini-batch gradient descent was decayed after training for every epoch to converge fast and prevent overfitting. Note that $\mathcal{E}$ here is defined slightly different from that in the main context, which contains both gene and cancer type embeddings.	}
	\label{alg:gene2vec}
\end{algorithm2e}

\clearpage

\subsection{Mathematical details of multi-head self-attention mechanism}
\label{sec:mathmatics}

For all SGA-affected genes $\left\{ g \right\}_{g=1}^m$ and the cancer type $s$ of a tumor $t$, we first mapped them to corresponding gene embeddings $\left\{ \mathbf{e}_g \right\}_{g=1}^m$ and a cancer type embedding $\mathbf{e}_s$ from a look-up table $\mathcal{E}\!=\!\left\{ \mathbf{e}_g\right\}_{g \in \mathcal{G}} \cap \left\{ \mathbf{e}_s \right\}_{s \in \mathcal{S}}$, where $\mathbf{e}_g$ and $\mathbf{e}_s$ are  real-valued vectors. From the implementation perspective, we treated cancer types in the same way as SGAs, except the attention weight of it is fixed to be ``1''.

The overall idea of producing the tumor embedding $\mathbf{e}_t$ is to use the weighted sum of cancer type embedding $\mathbf{e}_s$ and gene embeddings $\left\{ \mathbf{e}_g \right\}_{g=1}^m$ (\fref{fig:structure}b) :
\begin{equation}
\mathbf{e}_t = 1 \cdot \mathbf{e}_s + \sum\nolimits_{g} \alpha_g \cdot \mathbf{e}_g = 1 \cdot \mathbf{e}_s + \alpha_1 \cdot \mathbf{e}_1 + ... + \alpha_m \cdot \mathbf{e}_m.
\end{equation}

The attention weights $\left\{ \alpha_{g} \right\}_{g=1}^m$ are calculated by employing multi-head self-attention mechanism, using gene embeddings of SGAs $\left\{ \mathbf{e}_g \right\}_{g=1}^m$ in the tumor (\fref{fig:structure}c):
\begin{equation}
\alpha_1, \alpha_2, ..., \alpha_m = \text{Function}_{\text{Attention}}(\mathbf{e}_1, \mathbf{e}_2, ..., \mathbf{e}_m).
\label{eq:attention}
\end{equation}
The attention function $ \text{Function}_{\text{Attention}}$ is implemented as a sub-network.
In the case of single-head attention, there is only one single head parameter $\boldsymbol{\theta}_j$, and the unnormalized weights $\left\{ \beta_{g,j} \right\}_{g=1}^m$ can be derived as follows:
\begin{equation}
\beta_{g,j} = \boldsymbol{\theta}_j^\intercal \cdot \text{tanh}(W_0 \cdot \mathbf{e}_g), \ g = 1, 2, ..., m,
\label{eq:unnormalized}
\end{equation}
which are further normalized to single-head weights $\left\{ \alpha_{g,j} \right\}_{g=1}^m$:
\begin{equation}
\alpha_{1,j}, \alpha_{2,j}, ..., \alpha_{m,j} = \text{softmax}(\beta_{1,j}, \beta_{2,j}, ..., \beta_{m,j}),
\label{eq:normalized}
\end{equation}
where softmax function is defined as : $\alpha_g\!=\!\exp \left( \beta_g \right)\!/\!\sum_{g'=1}^m \exp \left( \beta_{g'} \right)$.
In the case of multi-head attention, there exist $h$ different parameters $\Theta\!=\!\left\{ \boldsymbol{\theta}_j \right\}_{j=1}^h$. Then multiple attention weights of each gene embedding are generated following Eq.~(\ref{eq:unnormalized},\ref{eq:normalized}) and summed up to be the final multi-head attention weight:
\begin{equation}
\alpha_g = \sum\nolimits_{j=1}^h \alpha_{g,j} = \alpha_{g,1}+\alpha_{g,2}+...+\alpha_{g,h}, \ g=1,2,...,m.
\label{eq:summultihead}
\end{equation}

\clearpage
\subsection{Evaluation metrics  of gene embedding space}
\label{sec:evaluationembedding}

We designed two metrics for evaluating whether the gene embedding space is fair using the Gene Ontology (GO)\cite{go00}. We mainly concentrated on evaluating whether SGA-affected genes share GO annotations in the ``biological process'' domain, based on the assumption that genes involved in a common biological process will likely share common functional impact. The top 1,474 frequently altered genes (affected by SGAs for more than 150 times across all the tumors in the dataset) were used for evaluation, assuming that the gene embeddings of rare SGAs may not be well learned. 

\textbf{NN accuracy:} We first designed a metric called ``nearest neighborhood (NN) accuracy'' as a measure of functional similarity among genes sharing similar gene embedding. It is defined as the expectation of whether a pair of genes ($g,c$) that are nearest neighbors in the embedding space share at least one same GO term: 
\begin{equation}
\text{NN accuracy} = \E\nolimits_{\mathbf{e}_c \in \text{NN}(\mathbf{e}_g)} \left[ \mathbb{1} \left( \text{GO}(g) \cap \text{GO}(c) \neq \emptyset \right) \right],
\label{eq:go-nnaccuracy}
\end{equation}
where $\mathbb{1}(\text{statement})$ is the indicator function;
$\text{GO}(g)$ the set of GO terms assigned to gene $g$; $\text{NN}(\mathbf{e}_g)$ the set of nearest neighbors of $\mathbf{e}_g$.
The expectation $\E$ is approximated by iterating over all possible pairs of genes. The higher NN accuracy, the functionally similar genes are more close to each other in the embedding space. 

\textbf{GO enrichment:} Apart from the NN accuracy, which only reflects the functional similarities between two adjacent genes in embedding space, we also evaluated whether a cluster of genes close in an embedding space share GO annotations through  ``GO enrichment'', which is defined as: 
\begin{equation}
\text{enrichment} = \frac{
	\E_{\text{Clust}(\mathbf{e}_g) = \text{Clust}(\mathbf{e}_c)} \left[ \mathbb{1}(\text{GO}(g) \cap \text{GO}(c) \neq \emptyset) \right]
}{
	\E_{g, c \in \mathcal{G}} \left[ \mathbb{1}(\text{GO}(g) \cap \text{GO}(c) \neq \emptyset) \right]
},
\label{eq:go-enrichment}
\end{equation}
where $\text{Clust}(g)$ is the cluster that gene $g$ belongs to.
GO enrichment considers the functional similarities of genes that are close in the embedding space. The larger it is, the higher correlated are the GO functions and clusters (and it equals to 1 in random case).

\clearpage
\subsection{Performance of GIT on real and shuffled data}

\label{sec:shuffle}

\begin{figure}[htpb]
	\centerline{\includegraphics[width=0.60 \linewidth]{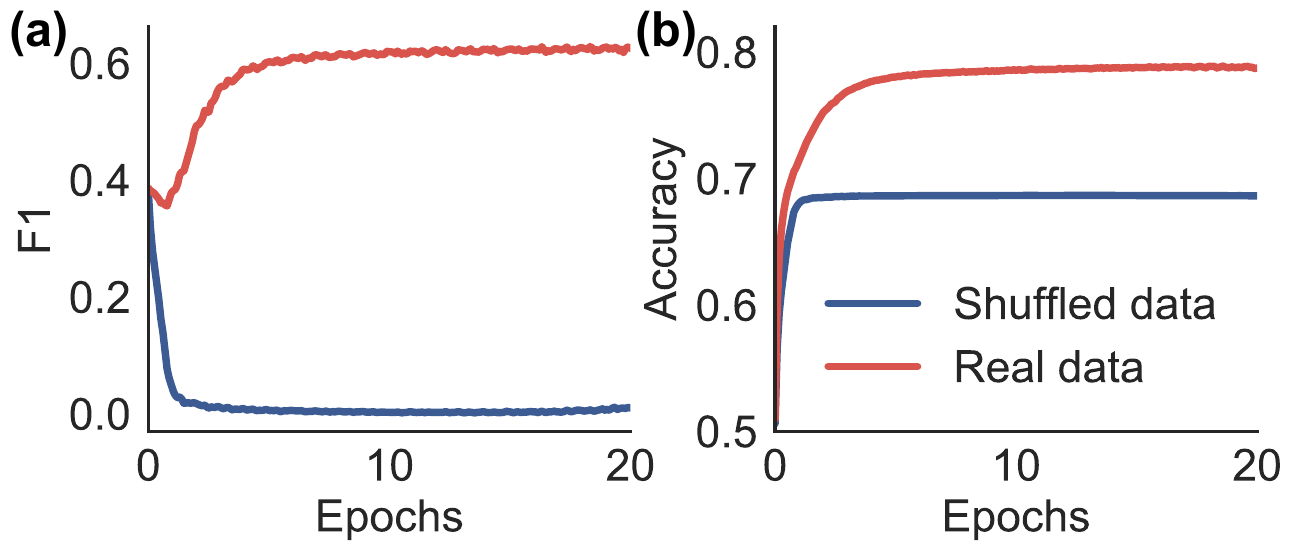}}
	\caption{
		\textbf{The change of F1 score and accuracy on the test set as GIT trains on real data or DEG-permuted data.}
	}
	\label{fig:shuffle}
\end{figure}

We plotted both F1 score and accuracy on the test set as the function of trained epochs  (Figure~\ref{fig:shuffle} ``real data''), which indicate that the model gains the capability of predicting DEGs as training proceeds, and finally reaches a stable state.

In order to validate that GIT is able to extract real statistical relationships between SGAs and DEGs, we randomly shuffled the positions of DEGs in the DEG vector of a tumor, i.e., randomly relabel DEG names, and then trained a GIT to predict DEGs from SGAs. We compared the performance of models trained with random datasets, by plotting F1 score and accuracy during the training of the models (Figure~\ref{fig:shuffle} ``shuffled data''). Note that, since most DEGs in the data are zeros, a trivial solution is to call every DEG as 0, which can also achieve good overall accuracy and minimize loss, but that will result in a low F1 because of low recall. Indeed, the test F1 score in the DEG-permutation case drops to a very low value due to the same reason.

\clearpage

\subsection{Enriched functions of gene clusters}

\label{sec:clusterfunction}

See \tref{tab:go} for the enriched functions of gene clusters. Fisher's exact test with Bonferroni correction (\textit{p}-value$<$0.05) was implemented on genes that belong to 40 clusters. 12 clusters of genes show to be significantly involved in at least one biological process. The genes in cluster 14, referred to as ``IFN pathway'', was further analyzed as a case study in \sref{sec:sga}, which is involved in viral defense response, immune response and cell surface signaling.

\begin{table}[h]
	\tbl{\textbf{Enriched gene ontologies in the  ``biological process'' domain of human beings (\textit{Homo sapiens}).}
		}
	{\scriptsize
		\begin{tabular}{ m{0.14\linewidth} m{0.18\linewidth} m{0.5\linewidth}  m{0.08\linewidth}  }\hline
			Cluster ID & Enriched gene ontology & Enriched biological process & \textit{p}-value \\
			\hline 
			2 & GO:0038003 & Opioid receptor signaling pathway & 2.09e-02 \\
			3 & GO:0050911 & Detection of chemical stimulus involved in sensory perception of smell & 3.16e-31 \\
			& GO:0007186 & G protein-coupled receptor signaling pathway & 5.27e-21 \\
			4 & GO:0007156 & Homophilic cell adhesion via plasma membrane adhesion molecules & 5.26e-03 \\
			& GO:0001568 & Blood vessel development & 4.64e-02 \\
			& GO:0048666 & Neuron development & 4.02e-02 \\
			& GO:0009653 & Anatomical structure morphogenesis & 3.17e-04 \\
			8 & GO:0045995 & Regulation of embryonic development & 3.77e-02 \\
			& GO:0007155 & Cell adhesion & 4.28e-02 \\
			14 (IFN pathway) & GO:0033141 & Positive regulation of peptidyl-serine phosphorylation of stat protein & 9.07e-30 \\
			& GO:0002323 & Natural killer cell activation involved in immune response & 6.02e-29 \\
			& GO:0042100 & B cell proliferation & 1.65e-26 \\
			& GO:0043330 & Response to exogenous dsrna & 1.05e-25 \\
			& GO:0002286 & T cell activation involved in immune response & 2.22e-24 \\
			& GO:0060337 & Type i interferon signaling pathway & 2.93e-21 \\
			& GO:0030183 & B cell differentiation & 1.40e-21 \\
			& GO:0051607 & Defense response to virus & 2.85e-18 \\
			& GO:0007596 & Blood coagulation & 5.77e-14 \\
			& GO:0006959 & Humoral immune response & 4.82e-15 \\
			& GO:0002250 & Adaptive immune response & 1.61e-12 \\
			& GO:0010469 & Regulation of signaling receptor activity & 2.83e-12 \\
			16 & GO:0050727 & Regulation of inflammatory response & 4.13e-02 \\
			23 & GO:0003272 & Endocardial cushion formation & 3.61e-02 \\
			& GO:0003179 & Heart valve morphogenesis & 9.79e-03 \\
			& GO:0007156 & Homophilic cell adhesion via plasma membrane adhesion molecules & 4.87e-02 \\
			& GO:0035295 & Tube development & 2.32e-03 \\
			& GO:0051960 & Regulation of nervous system development & 4.73e-02 \\
			& GO:0007399 & Nervous system development & 1.88e-02 \\
			25 & GO:0051179 & Localization & 1.19e-02 \\
			30 & GO:0000904 & Cell morphogenesis involved in differentiation & 4.15e-02 \\
			& GO:0007155 & Cell adhesion & 2.93e-02 \\
			& GO:0007275 & Multicellular organism development & 2.21e-03 \\
			35 & GO:0007156 & Homophilic cell adhesion via plasma membrane adhesion molecules & 1.60e-09 \\
			36 & GO:0040011 & Locomotion & 3.42e-02 \\
			40 & GO:0035589 & G protein-coupled purinergic nucleotide receptor signaling pathway & 4.50e-03 \\
			\hline
	\end{tabular}}\label{tab:go}
\end{table}

\clearpage

\subsection{Top genes by attention mechanism and mutation rates}
\label{sec:rank}

See \tref{tab:attentionall} for the full list of top 100 genes that are assigned by the attention mechanism. \Tref{tab:frequency} shows the top 5 genes that are most frequently mutated in both pan-cancer and single cancer types. It serves as the control group, in comparison to the attention mechanism results (\tref{tab:attention},\ref{tab:attentionall}; experiment group).

\begin{table}[h]
	\tbl{\textbf{List of candidate drivers whose corresponding SGAs have top 100 highest attention weights.}
		Boldfaced genes are known drivers according to TumorPortal\cite{sna14} and IntOGen\cite{intogen13} database.}
	{\scriptsize
		\begin{tabular}{
				m{0.06\linewidth} m{0.14\linewidth} 
				m{0.06\linewidth} m{0.14\linewidth} 
				m{0.06\linewidth} m{0.14\linewidth} 
				m{0.06\linewidth} m{0.14\linewidth} 
			}
			\hline
			Rank & Gene & Rank & Gene & Rank & Gene & Rank & Gene \\
			\hline 
			1 & \textit{\textbf{TP53}} & 26 & \textit{MUC5B} & 51 & \textit{KRTAP4-11} & 76 & \textit{CNTNAP3B} \\
			2 & \textit{\textbf{PIK3CA}} & 27 & \textit{LMTK3} & 52 & \textit{CYP4F11} & 77 & \textit{NKRF} \\
			3 & \textit{\textbf{RB1}} & 28 & \textit{\textbf{AHNAK}} & 53 & \textit{EP400} & 78 & \textit{\textbf{SETD2}} \\
			4 & \textit{\textbf{PBRM1}} & 29 & \textit{\textbf{VHL}} & 54 & \textit{\textbf{XRN1}} & 79 & \textit{\textbf{LAMA2}} \\
			5 & \textit{\textbf{PTEN}} & 30 & \textit{\textbf{FGFR3}} & 55 & \textit{MBD6} & 80 & \textit{AARS} \\
			6 & \textit{\textbf{CDH1}} & 31 & \textit{PHF20} & 56 & \textit{AR} & 81 & \textit{SPON1} \\
			7 & \textit{\textbf{CASP8}} & 32 & \textit{\textbf{STK11}} & 57 & \textit{ANKRD30BP2} & 82 & \textit{WRN} \\
			8 & \textit{\textbf{KRAS}} & 33 & \textit{OCA2} & 58 & \textit{PRICKLE2} & 83 & \textit{LHX1} \\
			9 & \textit{SLC1A6} & 34 & \textit{\textbf{GATA3}} & 59 & \textit{RGAG1} & 84 & \textit{\textbf{STAG2}} \\
			10 & \textit{POMC} & 35 & \textit{PCNX} & 60 & \textit{KRT23} & 85 & \textit{KSR1} \\
			11 & \textit{RRN3P2} & 36 & \textit{KRTAP4-9} & 61 & \textit{UGT1A1} & 86 & \textit{GCDH} \\
			12 & \textit{TFAM} & 37 & \textit{LRRIQ3} & 62 & \textit{PARP8} & 87 & \textit{E2F3} \\
			13 & \textit{CD163} & 38 & \textit{MRGPRF} & 63 & \textit{TMPRSS6} & 88 & \textit{PDHX} \\
			14 & \textit{WDFY3} & 39 & \textit{\textbf{HSP90AA1}} & 64 & \textit{FMN2} & 89 & \textit{CLUH} \\
			15 & \textit{WDR44} & 40 & \textit{CNTN3} & 65 & \textit{\textbf{CDKN2A}} & 90 & \textit{PRICKLE4} \\
			16 & \textit{CYP51A1} & 41 & \textit{WNK3} & 66 & \textit{DIP2B} & 91 & \textit{GLUD2} \\
			17 & \textit{ADARB2} & 42 & \textit{PTPRD} & 67 & \textit{TBP} & 92 & \textit{CROCC} \\
			18 & \textit{C9orf53} & 43 & \textit{PCDHB16} & 68 & \textit{ZNF624} & 93 & \textit{\textbf{IDH1}} \\
			19 & \textit{\textbf{BAP1}} & 44 & \textit{RPLP0P2} & 69 & \textit{FEM1B} & 94 & \textit{GRIA1} \\
			20 & \textit{TMPRSS13} & 45 & \textit{COL6A1} & 70 & \textit{CDKN2B} & 95 & \textit{DLG5} \\
			21 & \textit{SV2C} & 46 & \textit{TTC39B} & 71 & \textit{PDE4D} & 96 & \textit{SMURF2P1} \\
			22 & \textit{MYCBP2} & 47 & \textit{\textbf{PGR}} & 72 & \textit{ISLR2} & 97 & \textit{CACNA1C} \\
			23 & \textit{\textbf{MED24}} & 48 & \textit{TBC1D4} & 73 & \textit{FLRT3} & 98 & \textit{KIAA1377} \\
			24 & \textit{\textbf{CYLD}} & 49 & \textit{ANKRD36C} & 74 & \textit{ZFAT} & 99 & \textit{PTPRZ1} \\
			25 & \textit{CYLC2} & 50 & \textit{GPATCH8} & 75 & \textit{\textbf{SMARCA4}} & 100 & \textit{\textbf{PCSK5}} \\
			\hline
	\end{tabular}}\label{tab:attentionall}
\end{table}

\begin{table}[h]
	\tbl{\textbf{Top five SGA-affected genes for Pan-Cancer and a few selected cancer types, ranked according to alteration frequency, as the control group to GIT.}
		The corresponding experiment group, which is the selected candidate drivers of GIT model, is shown in Table~\ref{tab:attention}.
		The known cancer drivers according to TumorPortal\cite{sna14} and IntOGen\cite{intogen13} are marked in bold font.}
	{\begin{tabular}{m{0.05\linewidth}
				m{0.10\linewidth} 
				m{0.10\linewidth} 
				m{0.10\linewidth} 
				m{0.10\linewidth} 
				m{0.10\linewidth} 
				m{0.10\linewidth} 
			}
			\hline
			Rank & \multicolumn{1}{c}{PANCAN}  & \multicolumn{1}{c}{BRCA} & \multicolumn{1}{c}{HNSC} & \multicolumn{1}{c}{LUAD} & \multicolumn{1}{c}{GBM} & \multicolumn{1}{c}{BLCA}\\
			\hline
			1     &  \textit{\textbf{TP53}}   & \textit{\textbf{TP53}}  & \textit{\textbf{TP53}}   & \textit{TTN}  & \textit{CDKN2A}  & \textit{TTN} \\ 
			2   & \textit{TTN}    & \textit{\textbf{PIK3CA}}   & \textit{\textbf{CDKN2A}}  & \textit{\textbf{TP53}}  & \textit{CDKN2B}  & \textit{\textbf{TP53}} \\
			3     &  \textit{\textbf{PIK3CA}}  & \textit{TTN}  & \textit{TTN}   & \textit{CSMD3}  & \textit{C9orf53} & \textit{\textbf{ARID1A}} \\ 
			4    &  \textit{CSMD3}    & \textit{POU5F1B}  & \textit{\textbf{PIK3CA}}  & \textit{PCDHAC2}  & \textit{\textbf{EGFR}}  & \textit{DNAH5} \\ 
			5   &  \textit{MUC4}   & \textit{TRPS1}  & \textit{LINC00969} & \textit{MUC16}   & \textit{MTAP}  & \textit{\textbf{CDKN2A}} \\
			\hline
	\end{tabular}}\label{tab:frequency}
\end{table}

\clearpage

\subsection{Survival analysis based on raw SGAs}

\label{sec:sgasurvival}

\begin{figure}[htbp]
	\centerline{\includegraphics[width=0.8 \linewidth]{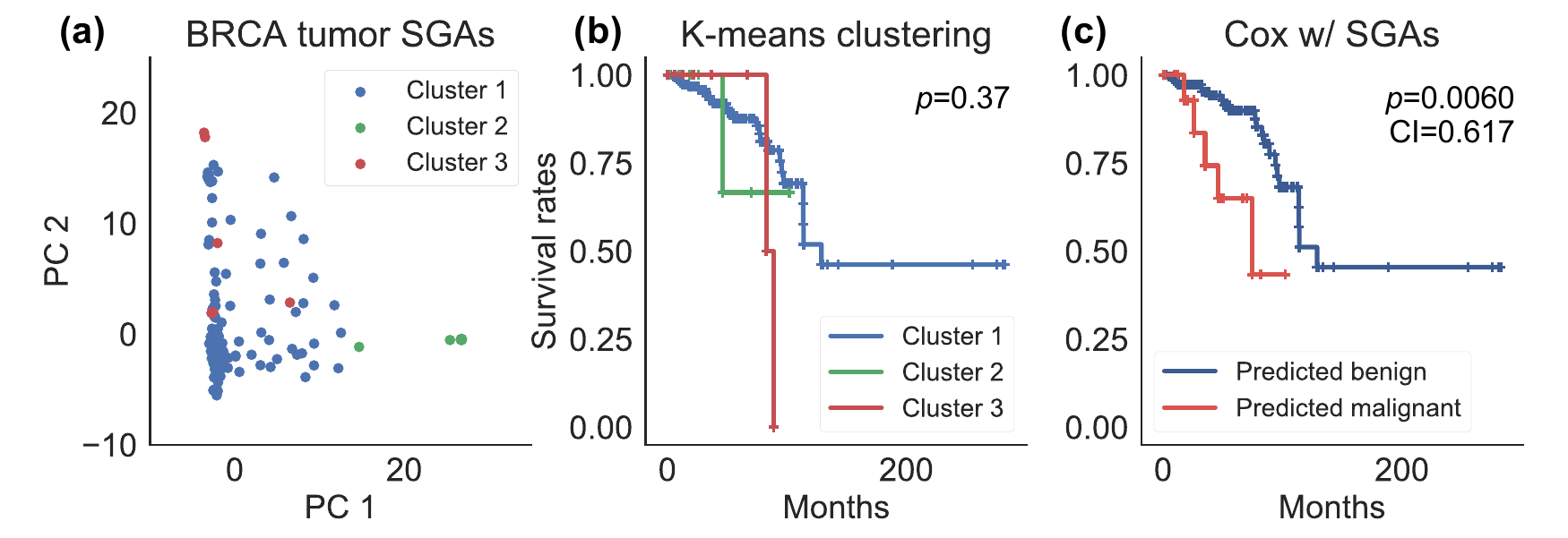}}
	\caption{
		\textbf{(a)} PCA plot showing \textit{k}-means clustering of BRCA tumors using their SGA vectors. Most tumors merge around the origin (Cluster 1; with a small number of SGAs), while others (Cluster 2,3; with a large number of SGAs) are outliers and far away from the origin.
		\textbf{(b)} KM estimators and log-rank test on the three BRCA tumor groups in the SGA space.
		\textbf{(c)} Cox regression using SGAs (top mutated 474 genes are used).
	}
	\label{fig:survival-sga}
\end{figure}

SGAs alone as tumor representations are not informative of predicting survival profiles. See \fref{fig:survival-sga} for survival analysis based on raw SGAs.

\end{document}